\documentclass[12pt,A4]{article}
\usepackage[english]{babel}
\usepackage{graphicx}

\newcommand{\alt}{\mathbin{\lower 3pt\hbox
   {$\rlap{\raise 5pt\hbox{$\char'074$}}\mathchar"7218$}}}
\newcommand{\agt}{\mathbin{\lower 3pt\hbox
   {$\rlap{\raise 5pt\hbox{$\char'076$}}\mathchar"7218$}}}

\begin{document}

\setcounter{footnote}{0}
\setcounter{equation}{0}
\setcounter{figure}{0}
\setcounter{table}{0}

\title{\large\bf Anderson transition in high dimension:
comments to arXiv:2403.01974 }

\author{\small  I. M. Suslov \\
\small P.L.Kapitza Institute for Physical Problems,
119334 Moscow, Russia \\
\small E-mail: suslov@kapitza.ras.ru\\
{}\\
\parbox{150mm}{\footnotesize \,	In the recent
submission arXiv:2403.01974, \mbox{Altshuler\,\,\,et\,\,\,al}
suggested a new approach to the Anderson transition in high
dimensions. The main idea consists in the use of the branching
graphs instead of high-dimensional lattices:  it does not look
very convincing, but we do not want  to stress this point. Since the
authors welcome comments, we put forward a lot of objections to
their exposition of the general situation.  The arising
hypothesis is given in the end.  } }

\date{}
\maketitle


\setcounter{footnote}{0}
\setcounter{equation}{0}
\setcounter{figure}{0}
\setcounter{table}{0}

In the recent submission, \mbox{Altshuler\,\,\,et\,\,\,al}
\cite{100} suggested a new approach to the Anderson
transition in high dimensions. The main idea, that
the branching graphs can be used instead of
high-dimensional lattices, does not look very convincing, but
we do not want to critisize it. There are a lot of
objections to their exposition of the general situation
(below $d$ is dimensionality of space, $\nu$ and $s$ are critical
exponents of the correlation length and conductivity, $g$ is
dimensionless conductance).

\vspace*{4mm}

 1. A disordered system with a Gaussian random potential can be
 {\it exactly} reduced to the $\phi^4$ field theory with a
 negative sign of the interaction constant \cite{1,2,3}. Such
 theory is non-renormalizable for $d>4$.  Renormalizability is
 analyzed on the diagrammatic level, when one deals with the
 usual impurity technique \cite{101,102}); so references to the
 "wrong" interaction or the replica trick are irrelevant. If a
 theory is non-renormalizable, then the ultraviolet cut-off (or
 the atomic scale) cannot be excluded from results. Consequently,
 the correlation length $\xi$ is not the only relevant length
 scale, and the single-parameter scaling \cite{103} becomes
 impossible.  Hence, $d=4$ is an upper critical
 dimension\,\footnote{\,The corresponding theory for
 $(4-\epsilon)$ dimensions was developed for a density of states
 \cite{4}, but not for conductivity. }:  it is a bare fact, which
 cannot be denied.
 \vspace*{4mm}

 2. Correspondence of a disordered system with any kind of the
 sigma-model is {\it approximate}. Sigma-models do not possess the
 upper critical dimension, and it can be clearly understood on
 the example of vector sigma-models. Fluctuations of the
 modulus of the vector order parameter are artificially
 suppressed in sigma-models, and it is well justified for
 $d=2+\epsilon$ \cite{104}. However, namely this fluctuation
 mode becomes catastrophically soft in approaching the upper
 critical dimension, and looks as a driving mechanism for its
 appearance.  It is evident from the Wilson theory \cite{1,105}.

\vspace*{4mm}

 3. Due to a failure with the upper critical dimension, the
 correspondence of the sigma-models with disordered systems is
 destroyed for $d>4$. Nevertheless, one can believe that such
 correspondence remains exact for $2<d<4$. However, it is only a
 belief. Alternatively, one can suggest, that a difference
 between sigma-models and disordered systems, being small for
 $d=2+\epsilon$,  is gradually increasing with space
 dimensionality\,\footnote{\,This possibility is strongly
 supported by the detailed analysis in the recent submission
\cite{200}. }. From this point of view, the Wegner
 high order corrections  \cite{106} can be related
 with this difference, and then they have nothing to do with
 disordered systems\,\footnote{\,In fact, Wegner himself
 discusses analogous possibilities \cite{107}.}.
 It removes the main argument against
 validity of the Vollhardt and W$\ddot o$lfle self-consistent theory
 \cite{108}; in contrast to sigma-models, this theory reproduces
 the upper critical dimension and gives a correct value for it.

\vspace*{4mm}

 4. The above conclusion is confimed by numerical results on
 multifractality \cite{9}, which are in a good agreement with the
 Wegner one-loop result \cite{109} (supported by self-consistent
 theory \cite{8}), and invite to ignore the high-order
 corrections.

 \vspace*{4mm}

 5. There exists a direct relation (see the end of \cite{9}
 or \cite{8}) between the high-order Wegner corrections and the
 high-gradient catastrophe \cite{107,110}.

 \vspace*{4mm}

 6. If one accepts hypothetically that $\nu=1/(d-2)$ is an exact
 result, then he will meet with essential problems concerning the
 dimensional regularization \cite{6}, which was used by Wegner.
 The accepted result is possible only if $\beta(g)=\epsilon-1/g$
 exactly, but such form of the $\beta$-function contradicts to
 the physical requirements in the small $g$ region \cite{103}.
 It looks that dimensional regularization is unable  to deal with
 such situation, while there are no problems for other
 regularizations, where all expansion coefficients depend on $d$.

\vspace*{4mm}

 7. All numerical results for $d>4$ are surely incorrect, since
 they are based on the single-parameter scaling. The Vollhardt
 and W$\ddot o$lfle  theory suggests a different  kind of
 scaling for high dimensions, and its implementation essentially
 change the results \cite{5}.

\vspace*{4mm}

 8. An accuracy of the result by Slevin and Ohtsuki ($\nu=1.57\pm
 0.02$ for $d=3$) should not be taken seriously, due to the
 evident problems in their treatment of scaling corrections
 \cite{12,13}. The rest of numerical results are not so
 categorical in rejection of $\nu=s=1$ for $d=3$.

\vspace*{4mm}

 9. In fact, all the raw numerical data (if they are taken for granted)
 can be reinterpreted in
 such way that they become compatible with the Vollhardt and
 W$\ddot o$lfle theory \cite{5}--\cite{11}: the key point is a
 structure of its corrections to scaling. Even if this theory is
 not exact, it suggests an example of the scaling picture, which
 cannot be rejected a priori. Correspondingly, the mentioned
 reinterpretation cannot be simply rejected. As a result, the
 error bars, given by numerical researchers, become unconvincing.

\vspace*{4mm}

 10. Suggestions by Garcia and others concerning the high
 dimensions are based on the poor logics, and cannot be
 considered as arguments.

\vspace*{4mm}

 11. A lot of physical experiments give $s=1$ for $d=3$
 \cite{14,15,16} and other confirmations of the self-consistent
 theory \cite{16,17}.

\vspace*{4mm}

 12. In fact, we believe that the Vollhardt and W$\ddot o$lfle theory is
 exact, since it can be justified without artificial assumptions
 \cite{18}.

\vspace*{4mm}

Looking at this and comparing with \cite{100}, one can come
to the following hypothesis:  the use of the branching graphs
corresponds to high-dimensional disordered systems,
which are treated  artificially  within a
single-parameter scaling, and described artificially  by the
nonlinear sigma-models. It looks as a formal analytical
continuation from a small vicinity of dimension $d=2$.

\end{document}